\documentclass[conference]{IEEEtran}

\usepackage[utf8]{inputenc}
\usepackage{url}
\usepackage{booktabs}
\usepackage{calc}
\usepackage{graphicx}
\usepackage{comment}
\usepackage{multirow}
\usepackage{algpseudocode,algorithm,algorithmicx}
\usepackage{comment}
\usepackage{paralist}
\usepackage{color}
\usepackage{paralist}
\usepackage[dvipsnames]{xcolor}
\usepackage{amsmath}
\usepackage[nocompress]{cite}
\usepackage{amssymb}
\usepackage{amsthm}
\usepackage{bm}
\usepackage{xspace}
\usepackage[inline]{enumitem}
\usepackage{csquotes}
\usepackage{graphicx}
\usepackage{soul,color}
\usepackage{tikz}                               
\usepackage{forest}
\usepackage[normalem]{ulem}

\usetikzlibrary{positioning}

\newcommand{\descr}[1]{\vspace{0.02in}\underline{{\em #1}}:}
\renewcommand{\paragraph}{\descr}

\makeatletter
\newcommand*{\rom}[1]{\expandafter\@slowromancap\romannumeral #1@}
\makeatother

\newcommand{\ie}{{i.e.,} }
\newcommand{\eg}{{e.g.,} }

\IEEEoverridecommandlockouts\IEEEpubid{\makebox[\columnwidth]{978-1-7281-1062-2/20/\$31.00 2020 $\copyright$ IEEE \hfill}\hspace{\columnsep}\makebox[\columnwidth]{ }}

\begin{document}
\title{UAVs Path Deviation Attacks:\\ Survey and Research Challenges}

\author{
	\IEEEauthorblockN{
		Francesco Betti Sorbelli\IEEEauthorrefmark{1},
		Mauro Conti\IEEEauthorrefmark{2},
		Cristina M.~Pinotti\IEEEauthorrefmark{1}, and
		Giulio Rigoni\IEEEauthorrefmark{3}
	}
	\IEEEauthorblockA{
		\IEEEauthorrefmark{1}Dept. of Comp. Sci. and Math., University of Perugia, Italy;
		email: \{francesco.bettisorbelli, cristina.pinotti\}@unipg.it
	}
	\IEEEauthorblockA{
		\IEEEauthorrefmark{2}Dept. of Comp. Sci. and Math., University of Padua, Italy;
		email: conti@math.unipd.it\\
	}
	\IEEEauthorblockA{
		\IEEEauthorrefmark{3}Dept. of Comp. Sci. and Math., University of Florence, Italy;
		email: giulio.rigoni@unifi.it\\
	}
}

\IEEEoverridecommandlockouts

\maketitle

\begin{abstract}
Recently, Unmanned Aerial Vehicles (UAVs) are employed for a plethora of civilian applications. 
Such flying vehicles can accomplish tasks under the pilot's eyesight within the range of a remote controller,
or autonomously according to a certain pre-loaded path configuration.
Different path deviation attacks can be performed by malicious users against UAVs.
We classify such attacks and the relative defenses based on the UAV's flight mode, i.e.,
(i) First Person View (FPV), 
(ii) civilian Global Navigation Satellite System based (GNSS), and 
(iii) GNSS ``plus'' auxiliary technologies (GNSS+), and on the multiplicity, i.e.,
(i) Single UAV, and
(ii) Multiple UAVs.
We found that
very little has been done to secure the FPV flight mode against path deviation.
In GNSS mode, spoofing is the most worrisome attack. 
The best defense against spoofing seems to be redundancy, such as adding vision chips to single UAV
or using multiple arranged UAVs.
No specific attacks and defenses have been found in literature for GNSS+
or for UAVs moving in group without a pre-ordered
arrangement. These aspects require further investigation.
\end{abstract}

\section{Introduction}
Unmanned Aerial Vehicles (UAVs), also known as drones, seem to have great potential in future civilian applications,
like precision agriculture~\cite{mogili2018review}, search and rescue~\cite{goodrich2008supporting}, 
localization~\cite{sorbelli2018range, sorbelli2019range}, etc.
Such potential is available not only for large companies but also for small businesses.
Thanks to their increasing availability, UAVs can be envisaged
for several applications~\cite{dorling2016vehicle}.
So far, the flight (with aircraft) was employed, for example, for delivering goods between hubs, 
and then ground vehicles transport the goods to consumers. 
Due to the involved costs, small companies are only confined to the use of ground vehicles. 
But now UAVs can extend the horizon also for small companies
allowing them to increase their business.

Other activities like roofs inspection, maintenance of pylons, or windows cleaning,
could be performed tomorrow by specialized UAVs.
Similarly, various activities in agriculture such as orchards pruning or pesticides spreading could be 
automated using  UAVs powered with high-precision eye-manual coordination.
So, smart UAVs are the next big revolution in UAV technology promising to provide new opportunities in different applications in terms of reduced risks and lower cost.

In all these activities
the most critical aspect is accurately flying through certain positions.
Even a small error in terms of UAV positioning made by the pilot or due to navigation satellite system
when harvesting fruits or cleaning windows can nullify the use of UAVs.
Positioning can be altered by external users for malicious purposes performing
specific attacks, such as GNSS spoofing which focuses on hijacking UAVs.
Under such an attack, the UAV cannot perform anymore the assigned tasks because
the path followed by UAVs is under the control of the attacker.
Also, attacks on UAVs guided by a human could be done by hacking the private ad-hoc local network
between the drone and its controller.

In this paper, 
we concentrate on the threats that can prevent
single or multiple UAVs to accomplish their path and/or reach the destination.
We primarily focus on commercial UAVs, available and affordable for public and small companies that have small sizes, limited battery lifetime, 
moderate payload capacity, and specific constraints in terms of computational power.
Despite their limited cost, such drones can be equipped with several external sensors (cameras, antennas) that facilitate communication with the external world, and single-board computers that extend their computational capability.  

Since we are interested in attacks that limit the flying ability of UAVs and their faculty of following predefined paths,
we organize  a {\em taxonomy}
that groups the attacks based on three different flight modes, i.e., First Person View (FPV), Global Navigation Satellite System based (GNSS),  and an enhanced GNSS, dubbed GNSS ``plus'' (GNSS+).
Then, 
we also sub-group the works based on the multiplicity of UAVs (\ie Single or Multiple).

\paragraph{Organization}
The rest of the paper is organized as follows.
Sec.~\ref{sec:related} reviews recent surveys on security aspects of UAVs.
Sec.~\ref{sec:taxonomy} introduces our taxonomy of the literature about path deviation attacks.
Sec.~\ref{sec:defenses} presents the defenses according to the proposed attack taxonomy.
Sec.~\ref{sec:future} explores future research directions.
Finally, Sec.~\ref{sec:conclusions} offers conclusions.

\section{Related Surveys in the Recent Past}\label{sec:related}
Recently, due to the increasing interest in UAVs,
several surveys on UAVs have been published in the literature
that, however, do not cover our focus.

In~\cite{fotouhi2019survey}, the authors provide a comprehensive study on UAV cellular communications.
Authors ponder on UAVs that can be used as flying Base Stations (BSs) extending and enhancing the user's capabilities especially for crowded areas where there is a lack of coverage.
They survey the issues, challenges, and opportunities of using UAVs for cellular communications with a little focus on safety and cyber physical security aspects.

The authors in~\cite{wang2019survey} survey UAV networks from a Cyber Physical System (CPS) perspective
dealing with security and privacy concerns on possible UAV networks.
Also, in~\cite{choudhary2018intrusion} security concerns are addressed in networked UAVs
presenting
Intrusion Detection Systems (IDSs).

In~\cite{krishna2017review}, 
the authors categorize the UAVs attacks according to three different attacks, i.e., 
{\em physical} on unattended and unsecured UAVs, 
{\em remote} exploiting back doors, bugs, or other UAV software vulnerabilities, 
and {\em target} aimed to a specific UAV's sensor, for example GPS or cameras.
Physical attacks are conduced by an intruder who finds
the UAV (or its ground control) left unattended and unsecured.
Given physical access to the UAV, the attacker can  inject malicious code on the UAV for further attacks
and transform the attack in
a remote one.
Target attacks are for a specific UAV's sensor, such as GPS or cameras.
The conclusion in that survey is that the first two categories of threats (physical and remote) 
are well investigated,
while the last category (target) has been ignored. 

Authors in~\cite{shakhatreh2019unmanned} present 
a cyber attack/challenges taxonomy in UAV systems based on three challenges, i.e.,
{\em confidentiality} that can be broken by malware injection or hijacking,
{\em integrity} that focuses on modifying information of UAV systems,
and {\em availability} that can be performed with the goal of flooding the communications links.

Authors in~\cite{schmidt2016survey} study the impact of Global Positioning System (GPS) spoofing attacks on different settings, like power distribution networks, ships, aircraft, trucks, train, and mobile phones.
It does not consider UAVs, though.
They conclude that spoofing is a nascent threat 
not yet trivially doable,
but with high potential to be severe if low-cost GPS spoofers will become common
and the economical value of the applications that use GPS continues to grow.
This last survey is on GPS spoofing and it is the one closest to our focus. 



\section{The proposed taxonomy}\label{sec:taxonomy}
In this section we introduce a taxonomy of the pertinent literature.
In Sec.~\ref{sec:tax-flight-modes} and~\ref{sec:tax-multiplicity}, we sketch the UAV's flight modes
and introduce the multiplicities.
In Sec.~\ref{sec:tax-attacks}, we describe the background on the attacks that deviate the UAV from its path.

\subsection{UAV's Flight Modes}\label{sec:tax-flight-modes}
The first taxonomy level represent the UAV's flight modes used to group the attacks.
We distinguish between three different UAV's flight modes:
\begin{itemize}
    \item \textbf{FPV}: UAVs are directly moved by a pilot's either eyesight or through a screen monitor using the remote controller. 
    
    \item \textbf{GNSS}: UAVs autonomously fly by retrieving the GNSS signals.
    They also rely on the Inertial Measurement Unit (IMU) which uses various sensors to control the flight.
    
    \item \textbf{GNSS+}: GNSS mode can be enhanced by adopting other technologies, 
    such as using images, WiFi, or cellular networks, to improve the flight robustness. 
    GNSS+ could be also GNSS enriched by a perception system.
\end{itemize}

\subsection{UAV's Multiplicity}\label{sec:tax-multiplicity}
In the second taxonomy level, we consider the multiplicity
\ie the number of UAVs that can take part together in a task:
\begin{itemize}
    \item \textbf{Single}: only a single UAV is deployed for a task.
   
    \item \textbf{Multiple}: many UAVs are deployed for a common task. 
\end{itemize}

Note that we consider any number of UAVs in FPV as a set of single UAV because a pilot is required for each UAV.
We use the multiplicity just to group the defenses.

\subsection{UAV's Attacks}\label{sec:tax-attacks}
We now describe the most significant attacks to deviate the UAV's path based on the flight mode.

\paragraph{Attacks in FPV}
In FPV flight mode, in-flight attacks can be substantially performed by hacking the local 
controller/UAV WiFi network. 
Nowadays, UAVs rely on WiFi links to which a remote controller, cameras, gimbals, and also
other smart devices like smartphones or tablets, can connect. 
Many manufacturers design proprietary protocols for those wireless communications, which in turn can be
highly more powerful than common wireless protocols in terms of performance.
The weakness of proprietary protocols is, however, their scarce ability to quickly respond to security threats~\cite{boulanger2005open}, 
letting the UAVs more susceptible to attacks of hackers.

\paragraph{Attacks in GNSS/GNSS+}
The GNSS and GNSS+ flight modes are based on the multiple interoperable constellations of satellites 
like GPS, GLONASS, BeiDou, Galileo, and more~\cite{xu2016gps,vagle2018multiantenna}.
To better understand the threats behind this flight mode, we  describe
how the GPS works~\cite{tippenhauer2011requirements}. 
The GPS uses a certain number of satellite transmitters $S_i$ 
located at known locations $L_i^S$ on their orbits. Each
transmitter is equipped with a synchronized clock 
and broadcasts at a certain time $t$ a carefully chosen 
navigation signal $s_i(t)$, which includes the spreading code, the phase, the satellite ID,
timestamps, and information on the satellites’ deviation from the
predicted trajectories. The signal propagates with light speed $c$.
The GPS receiver $V$ receives at time $t$ the combined signal $ g(L, t)$ of all the signals earlier broadcast
by the satellites in range:
$ g(L, t) = \sum_i {A_i s_i \left( t - {\| L_i^S - L\|}/{c} \right) + n(L, t)},$
where $A_i$ is the signal attenuation, $s_i ( t - {\| L_i^S - L\|}/{c} )$ is the  signal broadcast by $S_i$ at time $t - {\| L_i^S - L\|}/{c}$,
and $n(L, t)$ is the background noise. Note that,  
the signal $s_i$ has been encoded at the time 
the signal started from the satellite. That is,  the time $t$ the signal was received by the receiver  minus the time required to traverse the Euclidean distance 
$d_i=\| L_i^S - L\|$ between $S_i$ and $V$ at the light speed $c$.
Due to the properties of the signal coding, the receiver can extract 
from $g(L,t)$ all the single information $s_i(t)$. 
From there, the time delay ${\| L_i^S - L\|}/{c}$ is retrieved, and the {\em ranges} $d_i$ inferred.
In reality, the receiver which has a time offset $\delta$ infers the {\em pseudo-ranges} $r_i = d_i + \delta c$ because
the clock  of the receiver is not fully synchronized with that of the satellites.
With $d_i$ rewritten as the distance
$\sqrt{(x-x_i^S)^2+(y-y_i^S)^2+(z-z_i^S)^2}$
between the receiver  position $L = (x, y, z) \in \mathbb{R}^3$  and the satellite position $L_i^S = (x_i^S, y_i^S, z_i^S)$, $L$ and the offset $\Delta = c \cdot \delta$ 
can be determined by solving a system with at least four pseudo-ranges.
We say a receiver has a {\em lock} to a specific transmitter when it is already receiving data from that satellite. 
So, the receiver needs at least four locks to localize itself.

GPS receivers are vulnerable to different attacks
which depend on how the attacker modifies the satellites signal.
The attacker can easily deceive the receiver at any location $L'$
if it has full control over the input of such antenna. 
On civilian GPS receivers, the attacker is free to choose its position,
the delay, and the  satellite position.
The attacker can  claim a fake satellite location
because civilian GPS signals are not authenticated. Given
the right hardware, anyone can transmit his own GPS signals and their data content are not
checked for plausibility or consistency.
On military GPS receivers, instead, the attacker cannot change the data content of the GPS messages.

%


Common attacks on every GNSS are  blocking, jamming, and spoofing. 
GNSS jamming and blocking are techniques which create strong interference when detecting the GNSS 
signal with the goal of preventing the lock of the receiver to a satellite~\cite{warner2003gps}.
GNSS spoofing focuses on deceiving the legitimate GNSS signal and hence to corrupt the pseudo-ranges.
A GNSS spoofing attack 
requires the attacker to make the victim stop receiving signals from the legitimate satellites and start receiving the attacker’s
signals. 
GNSS spoofing can either be classified as {\em hard} or  
{\em soft} spoofing attack~\cite{noh2019tractor}.
The hard attack consists in the transmission of a very powerful GNSS signal, 
that can be considered as jamming attack at first, 
forcing the receiver to drop the original signal in favor of the fake one.
Once the victim has locked the new spoofed GNSS signal, 
the attacker can easily modify the GNSS signal as desired.
More complex is the soft attack, due to the fact that the signal strength of the spoofed signal 
has to be coherent to the legitimate one.
Once the victim locks into the new spoofed signal, the attacker can smoothly increase 
the strength of the spoofed signal, and finally move the target accordingly.

Meaconing is the simplest form of GNSS spoofing based on interception and rebroadcast of GNSS signals. 
It is a kind of hard spoofing.
The fake signals are rebroadcast after a delay, typically, 
with power higher than the original one, to confuse victims~\cite{schmidt2016survey}.
The Lift-of-Align (LoA)  attack sets the spoofing signal perfectly aligned to the legitimate signal, 
and then gradually increases its power so as the receiver will select the fake signal. 
LoA is a kind of soft spoofing~\cite{xu2018performance}.


We are not aware of specific attacks for GNSS+. 
A specific attack for GNSS+ should involve both GNSS and the additional guidance system used. 
We instead found some defenses that exploit the redundancy in the guidance systems.

\section{Defenses by Flight Mode and Multiplicity}\label{sec:defenses}
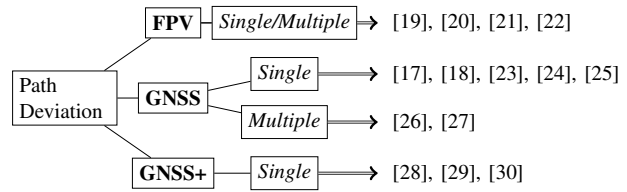
\begin{figure}[ht]
	\scalebox{0.8}{
	\begin{tikzpicture}	
		\node[draw, text=black, minimum size=0.5cm, text width=1.5cm] (A) at (0,0) {Path\\Deviation};
		
		\node[draw, text=black, minimum size=0.5cm] (A1) at (1.8,1.25) {\textbf{FPV}};
		\node[draw, text=black, minimum size=0.5cm] (A2) at (1.8,0) {\textbf{GNSS}};
		\node[draw, text=black, minimum size=0.5cm] (A3) at (1.8,-1.25) {\textbf{GNSS+}};
		
		\node[draw, text=black, minimum size=0.5cm] (A10) at (3.65,1.25) {\textit{Single/Multiple}};
		
		\node[draw, text=black, minimum size=0.5cm] (A21) at (3.65,0.4) {\textit{Single}};
		\node[draw, text=black, minimum size=0.5cm] (A22) at (3.65,-0.4) {\textit{Multiple}};
		
		\node[draw, text=black, minimum size=0.5cm] (A31) at (3.65,-1.25) {\textit{Single}};

		\draw (A) -- (A1);
		\draw (A) -- (A2);
		\draw (A) -- (A3);
		
		\draw (A1) -- (A10);
		
		\draw (A2) -- (A21);
		\draw (A2) -- (A22);
		
		\draw (A3) -- (A31);

		\draw (A1) -- (A10);
		
		\node[text=black, minimum size=0.5cm, anchor=west] (A10x) at (5.2,1.25) {\cite{shin2015security, pleban2014hacking, westerlund2019drone, dey2018security}};
		
		\node[text=black, minimum size=0.5cm, anchor=west] (A21x) at (5.2,0.4) {\cite{khalajmehrabadi2018real, 
		sun2018moving, liu2019analysis, xu2018performance, noh2019tractor}};
		\node[text=black, minimum size=0.5cm, anchor=west] (A22x) at (5.2,-0.4) {\cite{renyu2018spoofing, eldosouky2019drones}};
		
		\node[text=black, minimum size=0.5cm, anchor=west] (A31x) at (5.2,-1.25) {\cite{qiao2017vision, varshosaz2020spoofing,melikhova2018optimum}};
		
		\draw [double,->] (A10) -- (A10x);
		
		\draw [double,->] (A21) -- (A21x);
		\draw [double,->] (A22) -- (A22x);
		
		\draw [double,->] (A31) -- (A31x);
	\end{tikzpicture}
	}
	\caption{The proposed taxonomy of the inherent literature.}
	\label{fig:taxnomy}
\end{figure}

In this section, we discuss the defenses, if any, reported in the literature
for the attacks illustrated before.  
%
We organize the surveyed papers  as reported in Fig.~\ref{fig:taxnomy}.

\subsection{FPV Mode}\label{sec:FPV}
In FPV mode the UAV is guided by a pilot. 
The UAV can be controlled eyesight Line-of-Sight (LoS)
or remotely controlled through a camera mounted on the UAV and 
a built-in screen monitor on the controller.
The only available option for an attacker is directly taking control of the UAV by 
bypassing the connection between the UAV and the controller.
We found evidence in literature that attacks to this mode are feasible,
while almost no defenses were found, as listed below.

\subsubsection{Single/Multiple UAVs}
A work in~\cite{shin2015security} propose a system able to take control of criminal UAVs exploiting a radio frequency spoofing technique. 
The system emulates the original signals from the original controller. 
The downsize is that the UAV's model must be known beforehand 
to be able to emulate the signal correctly. 
The paper does not explain whether it is easy to extend the attack to other drone models. 

Authors in~\cite{pleban2014hacking} test the weakness of Parrot AR Drone 2.0.
As other UAVs, this model runs a Linux kernel and uses an open WiFi that leads to different attacks. 
Acquiring full access to the UAV over the unencrypted WiFi network is possible and not hard. 
The UAV creates an open hotspot, and because no password is required, 
authors were able to connect to it and scan its ports. 
Telnet port grants access to root shell with no password, 
and whereby to the entire UAV's operating system.  

Authors of~\cite{westerlund2019drone} show vulnerabilities 
able to alter the WiFi controller/UAV connection. For this task, they select two UAVs:
again the Parrot AR Drone 2.0 and Cheerson CX-10W Drone.
Those are toy-like UAVs sold everywhere, widely used, and ideal for various applications.
Attacks like Denial of Service (DoS), deauthentication, man-in-the-middle,
and packet spoofing are tested using very cheap hardware. 
Results show that all of those attacks were successful. 

Similarly, authors in~\cite{dey2018security} investigate security vulnerabilities associated with two more complex UAVs, i.e., DJI Phantom 4 Pro and Parrot Bebop 2. 
While Phantom 4 Pro is mainly susceptible to interception and jamming of the controller radio signal, Bebop 2 uses an open WiFi connection making it weak against various wireless attacks (\eg deauthentication attack, main-in-the-middle, etc).
Therefore both can be easily hijacked.
According to the authors, security improvement like WPA/WPA2 for encrypting the WiFi password may cause decrease in range and transmission speed.

In conclusion, the main issue disclosed in the FPV mode analysis is the lack of an encrypted WiFi connection. This means that not only the access to the WiFi is visible and open to everyone (because the password is not required) but also that all the network traffic between 
UAV and controller can be sniffed by any user employing simple techniques and using very affordable hardware.
So far, defenses in this field have not attracted great interest perhaps because the FPV UAVs have been mainly used  for recreational purposes. 
However, this flaw might be very dangerous.
There are many applications whose importance is increasing like trellis maintenance, windows cleaning, and such, in which an hijacked FPV UAV can do catastrophic damages. 
Therefore, the first defense is to use encrypted WiFi connections as long as they can be supported in terms of energy and computational power by the UAV's controller~\cite{shin2015security,dey2018security}.
Investigating the scenario of exchanging info between FPV UAV and Internet of Things (IoT) devices on the ground, seems worthy too.

\subsection{GNSS Mode}\label{sec:GNSS}
In this section, we consider the defenses for
preventing spoofing.

\subsubsection{Single UAV}
In the literature, GNSS attacks mainly focus on different targets  of UAVs.
However, UAVs can incur on the same attacks 
because their GNSS receivers are standard.
The only paper we are aware to conduct a GNSS spoofing attack to UAVs is~\cite{noh2019tractor}. 
The authors perform GNSS spoofing for a {\em safe hijacking attack} to move away a (terrorist) UAV from protected areas instead of incapacitating it.
They consider Phantom DJIs, Parrot Bebop 2, and 3DR Solo UAVs.
Simple hard spoofing is enough for hijacking DJIs and Parrot,
while for 3DR, based on ArduPilot, soft spoofing is required.
They successfully conduct the attack to all the three drones.

As for the attacks, there are not  ad-hoc defenses for single UAV.
Proposed defenses  for other targets
can be used for UAVs  
as long as they do not demand specific hardware or high computational burden.
As said, not all the defenses can be adapted.
An interesting case of countermeasure that cannot be moved to UAVs is that in~\cite{tippenhauer2011requirements}. 
Such countermeasure is based on the use of multiple  
receivers installed on the same target with a certain spacing. 
The authors exemplify their countermeasure on a cargo ship in which there is enough space to put several receivers apart one from the others. 
Clearly, such defense is  unfeasible on a single small UAV. 

\paragraph{GNSS Signal Property Technique}
We report here a 
set of papers that leverage the incoherence of the received counterfeit GNSS signals to defenses.

The authors in~\cite{khalajmehrabadi2018real} introduce and evaluate the
Time Synchronization Attack Rejection and Mitigation (TSARM) technique for time synchronization attacks over common GNSS receivers. 
This technique works in two steps: 
first, a dynamic model is introduced which analytically detects and models the GNSS attacks in the receiver's clock bias and drift, 
and second, the spoofed signatures, i.e., the estimated clock bias and drift, are accordingly modified based on the estimated attack.
So, the attacked GNSS receiver can ignore the inconsistent signatures and proceed with its normal operation applying adequate time corrections on the application.


Authors in~\cite{sun2018moving} 
exploit Signal Quality Monitoring (SQM) techniques for detecting spoofing attacks.
Originally designed for multipath detection,
SQM-based methods have good feasibility in a simpler context such as the
duel between spoofing and legitimate signals.
The authors improve the SQM technique proposing a new system that works by analyzing the initial interaction between spoofing and legitimate signals when the SQM metric fluctuates significantly.
Once both the fake and legitimate signals reach the victim,
the SQM fluctuation is empathized to gain the lock.
Such fluctuation causes an abnormality quantity of output values which allows taking adequate countermeasures.

These papers are based on sophisticated detection schemes
which, however, may become quickly obsolete
because powerful attackers may use advanced techniques  (\eg based on reinforcement algorithms, machine learning) for breaking the defenses.

\paragraph{IMU} 
The IMU-based defenses are countermeasures that exploit the redundancy of available knowledge at the UAV. 
IMU essentially grants another pool of information to which turn under dubious conditions.
The measurements from the IMU are fed to the Inertial Navigation System (INS) that helps the UAV to stay on track. It works using Kalman Filters (KFs) which elaborate a series of observed measurements over time to estimate unknown variables. 
Usually, KFs estimate the position and velocity errors, gyroscope bias, random noise errors, and accelerometer bias error.

The authors in~\cite{liu2019analysis} propose a technique based on the KF innovation sequence (\ie the output prediction error)
for a robust spoofing detection method.
The weakness of traditional KF methods is that they validate the correlation between the measurements and
the output prediction errors only in specific snapshots.
In this way, KFs are able to detect only rough GNSS spoofing attacks, \ie when the measurement drift
introduced by the spoofer is very large.
Knowing this, an attacker can smoothly spoof a victim.
The core idea behind the defense in that paper is to use the aforementioned KFs on a longer time window
testing all the possible accumulated errors  along the time and not only those on a specific snapshot.
The authors simulate GNSS spoofing attacks and prove the effectiveness of their detection method.

Similarly, in~\cite{xu2018performance} authors study the effects on the KF of two different spoofing attacks: meaconing 
and LoA attacks. 
The authors propose a detection method of the fake signal based on IMU error compensations obtained by applying the KF.
Results show that in the presence of meaconing and LoA attacks, 
the compensation of errors greatly increases, probably due to route instability (\ie small forced movements) induced from the spoofing signal. 
This instability creates inconsistency between IMU readings and expected values calculated from the GNSS signal, that leads to worthy compensations by KFs. 
In conclusion, the authors claim that their proposed system works well when detecting both the attacks. 

\subsubsection{Multiple UAVs}
This section presents papers that propose countermeasures 
when multiple UAVs
either preserve an arrangement during the flight or not.

In~\cite{renyu2018spoofing}, authors investigate a scenario in which an attacker with multiple antennas (for multiple spoofed signal) attacks a fleet of UAVs preserving the  fleet arrangement. 
First of all note that the attack can be only conducted using multiple antennas~\cite{tippenhauer2011requirements} because
a single attacker can spoof any number
of receivers in its transmission range, but they can be shifted only around the same coordinates $L'$.
Thus, 
the attacks to a fleet must be conducted  using multiple physical antennas.
Nonetheless, as demonstrated in~\cite{tippenhauer2011requirements,renyu2018spoofing},
there are a bounded number of physical positions from which the attack can be launched.
Generally, such attacker locations belong to a quadratic space, 
called hyperboloid, 
constrained by the pseudo-ranges between attacker and receivers and by the pseudo-ranges between claimed satellite and the two shifted positions. Not only, increasing the number of victims, the attacker locations belong to intersection
of hyperboloids, i.e., the possible locations are no longer squared space, but just set of points.
When the fleet has two UAVs, the possible locations are a set of hyperboloids.
When the fleet has three UAVs, the possible locations are a set of intersection of hyperboloids.
When there are four UAVs, the possible locations are a set of two points;
and from five or more  UAVs, the possible locations are a set of single points.
Therefore, it is particularly difficult to launch such an attack to arranged multiple drones.

Authors in~\cite{eldosouky2019drones} propose defenses against  multiple UAVs spoofing, not arranged.
The defense is based on cooperative localization.
Indeed, whenever a UAV doubts the authenticity of GNSS signal,
it can invoke a cooperative localization with other three  non-collinear UAVs in the fleet.
However, cooperative localization mechanism could not be reliable because other UAVs could be under attack.
To overcome this limitation, authors assume that only a single UAV at a time can be attacked.

From these results, we can state that the deployment of an arranged fleet of UAVs (that preserves the relative distances), instead of a single UAV, is enough to increase the defenses against GNSS spoofing attack.
But also just flying multiple UAVs together, without any kind of arrangement, offers opportunities of thwarting the attack
as illustrated in~\cite{eldosouky2019drones}. Multiplicity is a resource to be carefully exploited. 
However, one should also evaluate the overhead in accomplishing a complex task with multiple UAVs that cooperate for the same goal. 

\subsection{GNSS+ Mode}\label{sec:GNSS+}
Another simple but effective way to thwart spoofing attacks is to integrate a perception system in the UAV,
as it happens in GNSS+. 
We are not aware of papers that deal with attacks that simultaneously threat
the perception system and the UAV.
So, GNSS+ is more robust than GNSS.
The extra modules added to the UAVs are to be considered as a defense in themselves.

\subsubsection{Single UAV}
A commonly used additional technology on-board on UAVs is the combination of devices formed by gimbal and camera.
Gimbal allows the fixed camera to move along two or three axes enhancing the flight experience.

Authors in~\cite{qiao2017vision} propose a  vision-based approach using a camera. 
Basically, they combine together the velocity registered from the IMU and the calculated instant velocity from the Lucas-Kanade (LK) algorithm. 
The result is further compared with the GNSS velocity, and in case of discrepancies, a possible GNSS spoofing is detected.
The experiments are carried on a DJI Phantom 4, and the authors claim to be able to detect an attack, on average, between 5 seconds.

A camera for improving defenses against attacks is used in~\cite{varshosaz2020spoofing}. 
Authors substantially repeat  the same approach as~\cite{qiao2017vision}, 
but focusing on the UAV's path instead of the UAV's velocity.
From the camera images, using the Visual Odometry (VO) technique, 
the relative path is extracted and compared with the GNSS path. 
If the two paths do not fit each other, then the UAV is considered under a spoofing attack. 
To evaluate the similarities between the two measures, 
the authors propose different metrics that could detect spoofing attacks.

The system in~\cite{melikhova2018optimum} is based on Angle of Arrival (AoA) integrity-monitoring method. 
AoA method works on the assumption that all GNSS signals arrive from different positions and angles, while signals from a spoofer come all from the same direction.
The proposed solution adopts the generalized likelihood ratio test and an optimal decision-making algorithm is built to detect the presence of attacks using a three-antenna array (with a maximum distance between the antennas of half signal wavelength).
In a situation of possible attack in which the GNSS position cannot be trusted and thus considered unknown, results show that the detection is possible if the power level of the spoofed signals exceeds the power level of legitimate signals.

From these results, we can state that to flank GNSS receivers with other systems that help to control the route, as GNSS+ does, improves the system robustness.
Attack-oriented solutions that simultaneously compromise  the systems (GNSS, perception) involved in GNSS+
should be investigated. 

\section{Research Challenges and Future Works}\label{sec:future}
We conclude this work by reporting some lacks that attract our attention.
First of all, almost no defenses have been proposed for UAVs flying in FPV mode.
As the economical interest in first person drone activities increases, defenses
should be designed. 

Spoofing attacks still seem to be relatively easy
for single UAVs based on civilian GNSS receivers.
With the growth of UAV's applications and being easy to build
GNSS spoofers, GNSS spoofing  attacks can become more attractive
and profitable.
%
Several scheme to detect spoofing that operate on the check of signal properties have been devised, but their effectiveness can be reduced by highly capable attackers that carefully generate GNSS counterfeit signals to avoid triggering these detection schemes. 
We believe that the most effective way to improve resilience against spoofing attacks for single UAV is to strengthen the GNSS+ mode using different flanked perception systems.
Very interesting could be the integration of the 5G technology to UAVs, that will be soon available.
5G antennas will be an almost ready-to-use crowd-sourcing mechanism to cross-check the GNSS signal in the surroundings. 
%
Ad-hoc defenses for special UAV's tasks should also be investigated.
For example, in the case of a UAV patrolling an area (\eg repeating the same route constantly),
the success rate of an attack is higher due to the UAV predictability.
A defense mechanism based on multiple vision solutions that exploits the history of past patrols could be ideal.
Also, we have not found any paper that considers security on delivery with UAVs. This is
an area where applications daily bloom and attacks, like hijacking, could be extremely profitable.
Several algorithms for delivery have been proposed that integrate trucks and drones: this coupled delivery system could offer opportunities also for ensuring the delivery process.

We have also learned that it is substantially much more difficult to conduct a spoofing attack
to a fleet of UAVs that preserves an arrangement than to a single UAV.
However, accurately preserving the arrangement of multiple UAVs during the flight is not always possible. 
So, it could be interesting to verify up to inaccuracy of the arrangement (\ie relative distances among the drones), 
the attack is thwarted.
Also attacks for multiple drones that fly together without using GNSS system, which are now a hot-topic,
need attention.

Finally, attack and defense are two faces of the same coin; 
whenever new defensive mechanisms are introduced, new attacks can be carried on. Therefore, some future research topics could be attack-oriented to show  attacks on  GNSS+ mode
or on special drone applications.

\section{Conclusions}\label{sec:conclusions}
In this work, we reported the state-of-the-art of path deviation attacks against
different UAV's flight modes (FVP, GNSS, and GNSS+).
We learned  that
very little has been done to make secure the FPV flight mode.
In GNSS mode, spoofing is the most worrisome attack that can hijack the UAV.
The strong defenses seem to come from redundant devices
that assist the UAV in the flight.
An interesting future research direction is to consider ad-hoc solutions for thwarting attacks in
emergent UAVs applications, such as delivering.
UAV fleet that fly without GNSS is a new emergent trend not investigated at the moment.

\bibliographystyle{IEEEtran}
\bibliography{IEEEabrv,main-iauv-20}

\begin{thebibliography}{10}
\providecommand{\url}[1]{#1}
\csname url@samestyle\endcsname
\providecommand{\newblock}{\relax}
\providecommand{\bibinfo}[2]{#2}
\providecommand{\BIBentrySTDinterwordspacing}{\spaceskip=0pt\relax}
\providecommand{\BIBentryALTinterwordstretchfactor}{4}
\providecommand{\BIBentryALTinterwordspacing}{\spaceskip=\fontdimen2\font plus
\BIBentryALTinterwordstretchfactor\fontdimen3\font minus
  \fontdimen4\font\relax}
\providecommand{\BIBforeignlanguage}[2]{{%
\expandafter\ifx\csname l@#1\endcsname\relax
\typeout{** WARNING: IEEEtran.bst: No hyphenation pattern has been}%
\typeout{** loaded for the language `#1'. Using the pattern for}%
\typeout{** the default language instead.}%
\else
\language=\csname l@#1\endcsname
\fi
#2}}
\providecommand{\BIBdecl}{\relax}
\BIBdecl

\bibitem{mogili2018review}
U.~R. Mogili and B.~Deepak, ``Review on application of drone systems in
  precision agriculture,'' \emph{Procedia computer science}, vol. 133, pp.
  502--509, 2018.

\bibitem{goodrich2008supporting}
M.~A. Goodrich, B.~S. Morse, D.~Gerhardt, J.~L. Cooper, M.~Quigley, J.~A.
  Adams, and C.~Humphrey, ``Supporting wilderness search and rescue using a
  camera-equipped mini uav,'' \emph{Journal of Field Robotics}, vol.~25, no.
  1-2, pp. 89--110, 2008.

\bibitem{sorbelli2018range}
F.~B. Sorbelli, S.~K. Das, C.~M. Pinotti, and S.~Silvestri, ``Range based
  algorithms for precise localization of terrestrial objects using a drone,''
  \emph{Pervasive and Mobile Computing}, vol.~48, pp. 20--42, 2018.

\bibitem{sorbelli2019range}
F.~B. Sorbelli, C.~M. Pinotti, and V.~Ravelomanana, ``Range-free localization
  algorithm using a customary drone: Towards a realistic scenario,''
  \emph{Pervasive and Mobile Computing}, vol.~54, pp. 1--15, 2019.

\bibitem{dorling2016vehicle}
K.~Dorling, J.~Heinrichs, G.~G. Messier, and S.~Magierowski, ``Vehicle routing
  problems for drone delivery,'' \emph{IEEE Transactions on Systems, Man, and
  Cybernetics: Systems}, vol.~47, no.~1, pp. 70--85, 2016.

\bibitem{fotouhi2019survey}
A.~Fotouhi, H.~Qiang, and alias, ``Survey on uav cellular communications:
  Practical aspects, standardization advancements, regulation, and security
  challenges,'' \emph{IEEE Communications Surveys \& Tutorials}, vol.~21,
  no.~4, pp. 3417--3442, 2019.

\bibitem{wang2019survey}
H.~Wang, H.~Zhao, J.~Zhang, D.~Ma, J.~Li, and J.~Wei, ``Survey on unmanned
  aerial vehicle networks: A cyber physical system perspective,'' \emph{IEEE
  Communications Surveys \& Tutorials}, 2019.

\bibitem{choudhary2018intrusion}
G.~Choudhary, V.~Sharma, I.~You, K.~Yim, R.~Chen, and J.-H. Cho, ``Intrusion
  detection systems for networked unmanned aerial vehicles: a survey,'' in
  \emph{2018 14th International Wireless Communications \& Mobile Computing
  Conference (IWCMC)}.\hskip 1em plus 0.5em minus 0.4em\relax IEEE, 2018, pp.
  560--565.

\bibitem{krishna2017review}
C.~L. Krishna and R.~R. Murphy, ``A review on cybersecurity vulnerabilities for
  unmanned aerial vehicles,'' in \emph{2017 IEEE International Symposium on
  Safety, Security and Rescue Robotics (SSRR)}.\hskip 1em plus 0.5em minus
  0.4em\relax IEEE, 2017, pp. 194--199.

\bibitem{shakhatreh2019unmanned}
H.~Shakhatreh, A.~H. Sawalmeh, A.~Al-Fuqaha, and alias, ``Unmanned aerial
  vehicles (uavs): A survey on civil applications and key research
  challenges,'' \emph{IEEE Access}, vol.~7, pp. 48\,572--48\,634, 2019.

\bibitem{schmidt2016survey}
D.~Schmidt, K.~Radke, S.~Camtepe, E.~Foo, and M.~Ren, ``A survey and analysis
  of the gnss spoofing threat and countermeasures,'' \emph{ACM Computing
  Surveys (CSUR)}, vol.~48, no.~4, pp. 1--31, 2016.

\bibitem{boulanger2005open}
A.~Boulanger, ``Open-source versus proprietary software: Is one more reliable
  and secure than the other?'' \emph{IBM Systems Journal}, vol.~44, no.~2, pp.
  239--248, 2005.

\bibitem{xu2016gps}
G.~Xu and Y.~Xu, \emph{GPS: theory, algorithms and applications}.\hskip 1em
  plus 0.5em minus 0.4em\relax Springer, 2016.

\bibitem{vagle2018multiantenna}
N.~Vagle, A.~Broumandan, and G.~Lachapelle, ``Multiantenna gnss and inertial
  sensors/odometer coupling for robust vehicular navigation,'' \emph{IEEE
  Internet of Things Journal}, vol.~5, no.~6, pp. 4816--4828, 2018.

\bibitem{tippenhauer2011requirements}
N.~O. Tippenhauer, C.~P{\"o}pper, K.~B. Rasmussen, and S.~Capkun, ``On the
  requirements for successful gps spoofing attacks,'' in \emph{Proceedings of
  the 18th ACM conference on Computer and communications security}, 2011, pp.
  75--86.

\bibitem{warner2003gps}
J.~S. Warner and R.~G. Johnston, ``Gps spoofing countermeasures,''
  \emph{Homeland Security Journal}, vol.~25, no.~2, pp. 19--27, 2003.

\bibitem{noh2019tractor}
J.~Noh, Y.~Kwon, Y.~Son, H.~Shin, D.~Kim, J.~Choi, and Y.~Kim, ``Tractor beam:
  Safe-hijacking of consumer drones with adaptive gps spoofing,'' \emph{ACM
  Transactions on Privacy and Security (TOPS)}, vol.~22, no.~2, pp. 1--26,
  2019.

\bibitem{xu2018performance}
R.~Xu, M.~Ding, Y.~Qi, S.~Yue, and J.~Liu, ``Performance analysis of gnss/ins
  loosely coupled integration systems under spoofing attacks,'' \emph{Sensors},
  vol.~18, no.~12, p. 4108, 2018.

\bibitem{shin2015security}
M.~Donatti, F.~Frazatto, L.~Manera, T.~Teramoto, and E.~Neger, ``Radio
  frequency spoofing system to take over law-breaking drones,'' in \emph{2016
  IEEE MTT-S Latin America Microwave Conference (LAMC)}.\hskip 1em plus 0.5em
  minus 0.4em\relax IEEE, 2016, pp. 1--3.

\bibitem{pleban2014hacking}
J.-S. Pleban, R.~Band, and R.~Creutzburg, ``Hacking and securing the ar. drone
  2.0 quadcopter: investigations for improving the security of a toy,'' in
  \emph{Mobile Devices and Multimedia: Enabling Technologies, Algorithms, and
  Applications 2014}, vol. 9030, 2014, p. 90300L.

\bibitem{westerlund2019drone}
O.~Westerlund and R.~Asif, ``Drone hacking with raspberry-pi 3 and wifi
  pineapple: Security and privacy threats for the internet-of-things,'' in
  \emph{2019 1st International Conference on Unmanned Vehicle Systems-Oman
  (UVS)}.\hskip 1em plus 0.5em minus 0.4em\relax IEEE, 2019, pp. 1--10.

\bibitem{dey2018security}
V.~Dey, V.~Pudi, A.~Chattopadhyay, and Y.~Elovici, ``Security vulnerabilities
  of unmanned aerial vehicles and countermeasures: An experimental study,'' in
  \emph{2018 31st International Conference on VLSI Design (VLSID)}.\hskip 1em
  plus 0.5em minus 0.4em\relax IEEE, 2018, pp. 398--403.

\bibitem{khalajmehrabadi2018real}
A.~Khalajmehrabadi, N.~Gatsis, D.~Akopian, and A.~F. Taha, ``Real-time
  rejection and mitigation of time synchronization attacks on the global
  positioning system,'' \emph{IEEE Transactions on Industrial Electronics},
  vol.~65, no.~8, pp. 6425--6435, 2018.

\bibitem{sun2018moving}
C.~Sun, J.~W. Cheong, A.~G. Dempster, L.~Demicheli, E.~Cetin, H.~Zhao, and
  W.~Feng, ``Moving variance-based signal quality monitoring method for
  spoofing detection,'' \emph{GPS Solutions}, vol.~22, no.~3, p.~83, 2018.

\bibitem{liu2019analysis}
Y.~Liu, S.~Li, Q.~Fu, Z.~Liu, and Q.~Zhou, ``Analysis of kalman filter
  innovation-based gnss spoofing detection method for ins/gnss integrated
  navigation system,'' \emph{IEEE Sensors Journal}, vol.~19, no.~13, pp.
  5167--5178, 2019.

\bibitem{renyu2018spoofing}
Z.~Renyu, S.~C. Kiat, W.~Kai, and Z.~Heng, ``Spoofing attack of drone,'' in
  \emph{2018 IEEE 4th International Conference on Computer and Communications
  (ICCC)}.\hskip 1em plus 0.5em minus 0.4em\relax IEEE, 2018, pp. 1239--1246.

\bibitem{eldosouky2019drones}
A.~R. Eldosouky, A.~Ferdowsi, and W.~Saad, ``Drones in distress: A
  game-theoretic countermeasure for protecting uavs against gps spoofing,''
  \emph{IEEE Internet of Things Journal}, 2019.

\bibitem{qiao2017vision}
Y.~Qiao, Y.~Zhang, and X.~Du, ``A vision-based gps-spoofing detection method
  for small uavs,'' in \emph{13th International Conference on Computational
  Intelligence and Security (CIS)}.\hskip 1em plus 0.5em minus 0.4em\relax
  IEEE, 2017, pp. 312--316.

\bibitem{varshosaz2020spoofing}
M.~Varshosaz, A.~Afary, B.~Mojaradi, M.~Saadatseresht, and E.~Ghanbari~Parmehr,
  ``Spoofing detection of civilian uavs using visual odometry,''
  \emph{International Journal of Geo-Information}, vol.~9, no.~1, p.~6, 2020.

\bibitem{melikhova2018optimum}
A.~P. Melikhova and I.~A. Tsikin, ``Optimum array processing with unknown
  attitude parameters for gnss anti-spoofing integrity monitoring,'' in
  \emph{2018 41st International Conference on Telecommunications and Signal
  Processing (TSP)}.\hskip 1em plus 0.5em minus 0.4em\relax IEEE, 2018, pp.
  1--4.

\end{thebibliography}

\end{document}